\documentclass[conference]{IEEEtran}
\IEEEoverridecommandlockouts
\usepackage{cite}
\usepackage{amsmath,amssymb,amsfonts}
\usepackage{algorithmic}
\usepackage{graphicx}
\usepackage{textcomp}
\usepackage{xcolor}
\def\BibTeX{{\rm B\kern-.05em{\sc i\kern-.025em b}\kern-.08em
    T\kern-.1667em\lower.7ex\hbox{E}\kern-.125emX}}
\begin{document}

\title{SALF-MOS : Speaker Agnostic Latent Features Downsampled for MOS Prediction \\
}  

\author{\IEEEauthorblockN{Saurabh Agrawal, Raj Gohil, Gopal Kumar Agrawal, Vikram C M, Kushal Verma}
\IEEEauthorblockA{Samsung R\&D Institute Bangalore, India}
sauagr17@gmail.com, raj1996gohil@gmail.com, (gopal.kumar, vikram.cm, kushal.verma)@samsung.com}

\maketitle
\begin{abstract}
Speech quality assessment is a critical process in selecting text-to-speech synthesis (TTS) or voice conversion models. Evaluation of voice synthesis can be done using objective metrics or subjective metrics. Although there are many objective metrics like the Perceptual Evaluation of
Speech Quality (PESQ), Perceptual Objective Listening Quality Assessment (POLQA) or Short-Time Objective Intelligibility (STOI) but none of them is feasible in selecting the best model. On the other hand subjective metric like Mean Opinion Score is highly reliable but it requires a lot of manual efforts and are time-consuming. To counter the issues in MOS Evaluation, we have developed a novel model, Speaker Agnostic Latent Features (SALF)-Mean Opinion Score (MOS) which is a small-sized, end-to-end, highly generalized and scalable model for predicting MOS score on a scale of 5. We use the sequences of convolutions and stack them to get the latent features of the audio samples to get the best state-of-the-art results based on mean squared error (MSE), Linear Concordance Correlation coefficient (LCC), Spearman Rank Correlation Coefficient (SRCC) and Kendall Rank Correlation Coefficient (KTAU).\footnote{The paper is accepted at SPCOM 2024}
\end{abstract}


\section{\textbf{INTRODUCTION}}
For speech synthesis and voice conversion, the most reliable methods for evaluating voice is Mean Opinion Score (MOS) where different numbers of human evaluators are advised to evaluate the speech quality based on many parameters like voice consistency, pitch consistency, pronunciation consistency, voice naturalness, noise presence in voice and many other parameters depending on voice assessment. Although MOS is the most widely used method for evaluating Text-to-Speech (TTS) models but it has some challenges. Human participants need to listen to each audio sample and rate it on a 5-point scale, which requires a large amount of patience for listening to audios, rating them, and monetary funds to facilitate these listening tests are overheads. Generally, for each audio, a good number of listeners are required (around 20 to overcome score bias) who have the domain knowledge of the language. This can also serve as a challenge if the language is a low-resource language where the source of audio is limited and human evaluators are also limited. Listening to the audio and evaluating the MOS is time and resource consuming especially when the same test has to be repeated again and again for voice samples and hence it is not scalable. One also needs to control the listening environment and hardware specifications which further adds constraints to the MOS tests. 

In recent Self-Supervised Learning (SSL) mechanisms have been used to evaluate the MOS more precisely using deep learning mechanisms. In recent works, SSL representations like Wav2Vec~\cite{b13}, HuBERT~\cite{b14}, WavLM~\cite{b15}, TERA~\cite{b16} which have been trained on a large amount of data and the resultant high-quality speech representations are used for MOS prediction. Although several neural based approaches have been developed, there are still many challenges based on model parameters, generalization of one model over other data sets and general purpose prediction model.

\begin{figure}[htb]
    
  \centering
  \centerline{\includegraphics[width=8.5cm]{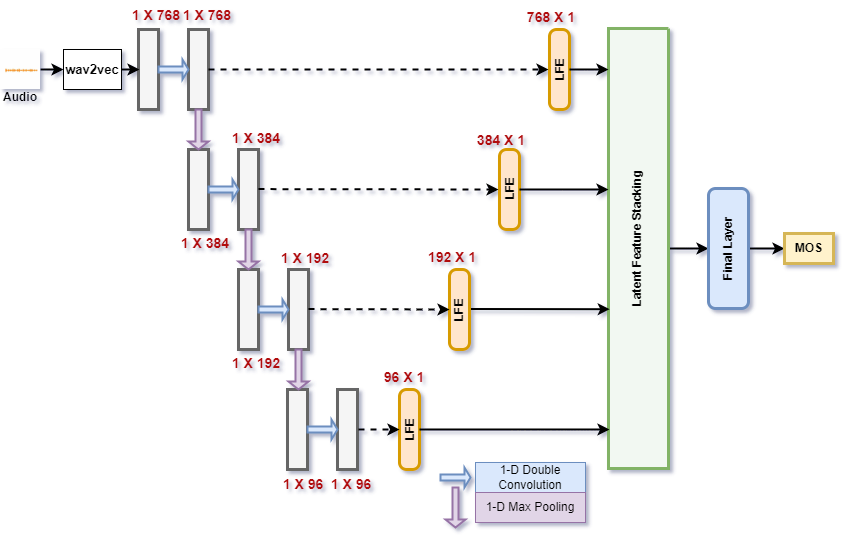}}
\caption{SALF-MOS Model Architecture}
\label{fig:res1}
\end{figure} 

In this paper, we will present Speaker Agnostic Latent Features Downsampled for MOS Prediction (SALF-MOS) which is shown in Fig.~\ref{fig:res1}. The proposed  model that generalizes well and sets new state-of-the-art (SOTA) results for evaluating TTS and Voice Conversion models. Before going over SALF-MOS lets explore some past works done for MOS prediction.

MOS prediction started with AutoMOS~\cite{b25} which uses Quality-Net~\cite{b26} model to predict Perceptual Evaluation of Speech Quality (PESQ)~\cite{b17}, a metric which compares a degraded speech signal with a reference speech signal. Work~\cite{b3} mentions training whisper (small) using feature extraction on source data and adapting the feature extracted on target data for MOS evaluation. Later, MOSA-Net~\cite{b27}, an improved version for AutoMOS was introduced. The MOSA-Net was gradually developed for cross domain model where multiple features including Mel Spectogram and SSL features were combined for MOS prediction.
Further MOSNET~\cite{b4} was introduced. MOSNET predicts MOS from magnitude spectrum using a CNN-BLSTM with a loss function which is a combination of frame level loss and utterance loss.
Since MOSNet uses the average MOS rating of each audio by different listeners, bias of each judge and variation will not be modelled in the right way. To encounter this issue, MBNet~\cite{b5} was proposed. MBNet leverages the mean and bias subnets to leverage the individual judge scores via two networks, namely  meanNet and biasNet networks.
MBNet removes biasNet at inference and only retains meanNet due to which model becomes inefficient. To counter this, an encoder-decoder architecture named LDNet~\cite{b6} was introduced to reduce the waste of parameters. After these kinds of model, combinations of SSL model started gaining attention. 
DDOS \cite{b8} which relies on pre-training of wav2vec2~\cite{b24} features on 2 modules - regression and distribution heads having Deep Neural Nets and Attentive Pooling mechanism for removing the domain mismatch from SSL model. 
UTMOS \cite{b9}, which is an ensemble learning of strong and weak learners where strong learners incorporates SSL model while weak learners incorporates Machine Learning models. It uses data Domain Id, Listener Id, phoneme sequence and reference sequence. 
FUSION-SSL \cite{b10} fuses the results of seven SSL models to predict MOS value.
MOSPC\cite{b7} uses seven SSL models - wav2vec (small, large, lv60),  hubert (base), wavlm (base, base+, large) for comparison of current audio in the data set with the next audio using C-mixup algorithm and helps the model to pay more attention to the current audio. 
NORESQA-MOS \cite{b11} uses non-matching references as a form of conditional to ground the MOS estimation by neural networks.

All these works heavily rely on combining various SSL models, pre-training and fine-tuning them on synthetic speech, dependent on listener id, dependent on domain id which serves as a major challenge for generalizing the model across different speakers. To encounter this issue we are introducing SALF-MOS, a Speaker Agnostic MOS Model which does not rely on fine-tuning SSL models, pre-training SSL model, Domain Id and Listener Id.
\begin{table*}
\centering
\caption{\centering Comparison of proposed model with the state of the art models for various data sets bench marked on metrics like MSE, LCC, SRCC, KTAU}
\label{table:results}
\resizebox{\textwidth}{!}{%
\begin{tabular}{|l|l|l|l|l|l|l|l|l|l|l|l|l|l|l|l|l|l|} 
\hline
            &                           & \multicolumn{16}{c|}{Dataset}                                                                                                    \\ 
\cline{3-18}
Model       & Size(No. of Parameters)   & \multicolumn{4}{c|}{BVCC}      & \multicolumn{4}{c|}{VCC2018}   & \multicolumn{4}{c|}{SOMOS}    & \multicolumn{4}{c|}{THMINTQI}  \\ 
\cline{3-18}
            &                           & MSE   & LCC    & SRCC  & KTAU  & MSE    & LCC   & SRCC  & KTAU  & MSE   & LCC   & SRCC  & KTAU  & MSE   & LCC   & SRCC  & KTAU   \\ 
\hline
\textbf{SALF-MOS}    & \multicolumn{1}{c|}{\textbf{1574}} & \textbf{0.144} & \textbf{0.948}  & \textbf{0.946} & \textbf{0.819}     & \textbf{0.336} & \textbf{0.825} & \textbf{0.829} & \textbf{0.678}     & \textbf{0.15}  & \textbf{0.773} & \textbf{0.771} & \textbf{0.583}     & \textbf{0.383}  & \textbf{0.795}  & \textbf{0.747}  & \textbf{0.576}      \\ 
\hline
NORESQA-MOS & \multicolumn{1}{c|}{92M}  & 0.17  & 0.89   & 0.87  & -     & -      & -     & -     & -     & -     & -     & -     & -     & -     & -     & -     & -      \\ 
\hline
UTMOS       &   \multicolumn{1}{c|}{-}  & 0.219 & 0.8822 & 0.88  & 0.707 & -      & -     & -     & -     & -     & -     & -     & -     & -     & -     & -     & -      \\ 
\hline
DDOS        & \multicolumn{1}{c|}{-} & 0.201 & 0.877  & 0.875 & 0.701 & -      & -     & -     & -     & -     & -     & -     & -     & -     & -     & -     & -      \\ 
\hline
MOSNet      & \multicolumn{1}{c|}{-} & 0.816 & 0.294  & 0.263 & -     & 0.538  & 0.643 & 0.589 & -     & 0.24  & 0.515 & 0.498 & 0.346 & -     & -     & -     & -      \\ 
\hline
LDNet       & \multicolumn{1}{c|}{0.96M} & 0.338 & 0.774  & 0.773 & 0.582 & 0.441  & 0.664 & 0.626 & 0.465 & 0.223 & 0.584 & 0.568 & 0.401 & -     & -     & -     & -      \\ 
\hline
MBNet       & \multicolumn{1}{c|}{1.38M} & 0.433 & 0.727  & 0.753 & 0564  & 0426   & 0.68  & 0.647 & -     & -     & -     & -     & -     & -     & -     & -     & -      \\ 
\hline
Fusion-SSL  & \multicolumn{1}{c|}{-} & 0.156 & 0.902  & 0.901 & 0.735 & 0.359  & 0.74  & 0.711 & 0.542 & -     & -     & -     & -     & -     & -     & -     & -      \\ 
\hline
MOSPC       & \multicolumn{1}{c|}{-} & 0.148 & 0.906  & 0.906 & 0.742 & 0.352  & 0.748 & 0.721 & 0.551 & -     & -     & -     & -     & -     & -     & -     & -      \\ 
\hline
LE-SSL-MOS  & \multicolumn{1}{c|}{-} & -     & -      & -     & -     & -      & -     & -     & -     & -     & -     & -     & -     & 0.951 & 0.537 & 0.517 & 0.388  \\
\hline
\end{tabular}%
}
\end{table*} 
\section{\textbf{PROPOSED METHOD}}
In this paper, we present SALF-MOS which is inspired from UNET~\cite{b12}. SALF-MOS is a fast, smaller-sized, end-to-end model designed for MOS prediction with down sampled features combined with linearly stacked networks to predict the MOS scores.

Although MOS is a subjective assessment but our introduced model can counter the other metrics like PESQ, POLQA and and STOI. Objective metrics for speech quality like PESQ~\cite{b17}, POLQA~\cite{b18} and STOI~\cite{b19} removes the dependency on speaker listening tests but have limited correlation with speech quality assessment for TTS and their application is limited to specific speech quality assessment which fails them for assessment of voice having background noise or voice transmission protocols or voice conversion systems. Since many features can be used to predict the MOS like MFCC, Mel Spectrogram, Magnitude Spectrogram and F0.  However, SSL features are gaining a lot of attention these days due to their generalization ability on large amounts of speech and language related tasks. One such SSL model, wav2vec~\cite{b13} encodes raw speech data via multiple convolutional layers to obtain abstract and latent representation. We used wav2vec, MFCC, LFCC and x-vector features for experimenting our model. Though, LFCC outperforms a few of the models while SSL based features outperform all other models and sets new state-of-the-art (SOTA) results for MOS prediction.

\subsection{\textbf{Model Architecture}}
Our Model consists of four Double Convolutions with three down sampling layers as mentioned in U-Net \cite{b12} which helps to extract latent features from the wav2vec. The system of four Double Convolutions and three down sampling is referred to a model of depth four. Each Double Convolution consists of 1-D Convolution with kernel size of 3, stride of 1 and padding of 1 followed by Batch Normalization over 1D with ReLU Activation. This process is followed again to make it a Double Convolution. The features extracted from Double Convolution are fed to linear layer for Latent Feature Extraction (LFE) and also down sampled using a kernel size of 2 and stride of 2. This down sampling reduces the dimension by a factor of 2 and is again followed by Double Convolution. Each feature extracted from double convolution from the model are then fed to linear layers which serve the purpose for extracting latent features. These features are stacked together to represent the feature mapping layer of various down sampled features. This stacked layer is sent to final linear layer which is used to learn the final representation of Mean Opinion Scores for the given audio.

Our model SALF-MOS introduces feature compression with smaller model size and removes the dependency of training SSL model. It also removes the dependency of multiple loss functions as used in ~\cite{b4, b5} and also removes dependency on combining several SSL models as in \cite{b7, b10} which increases the model size and introduces model training complexity. Our model is dynamic to various voices and pays more attention to correctly evaluate the quality of voice samples as compared to other model and generalizes well on various kinds of MOS data sets. Our model outperforms all other model in terms of MSE, LCC, SRCC and KTAU.

SALF-MOS consists of five main components:

1. Feature Generation using wav2vec

2. Latent Feature Compression using down sampling

3. Latent Feature Extraction using LFE layer

4. Latent Feature Stacking to learn the mapping of LFE with final layer

5. MOS Generation using final layer

These parts enhances our model performance to achieve SOTA results without any fine-tuning or pre-training of SSL models.

\section{\textbf{Experiments and Results}}
Our model generalizes for all the data sets shown in the Table~\ref{table:results} which consists of large and small number of data points. Our small sized model performs well with a varied number of audio samples. We experimented with a learning rate of 1e-4 and used L1 Loss and Stochastic Gradient Descent as the optimizer with a batch size of 4 per epoch. We used a single A10 GPU of 26GB on Ubuntu 20.04 with 4 cores of CPU each 16GB RAM operating at 3.00GHz. We used early stopping with 20 epochs to prevent model overfitting. Training, Validation and Testing data are splitted int the ratio of 8:1:1 and all results shown in this paper are for testing data.

\subsection{\textbf{Dataset}}\label{AA}
\begin{figure}[htb]
\begin{minipage}[b]{\linewidth}
    
  \centering
  \centerline{\includegraphics[width=8.5cm]{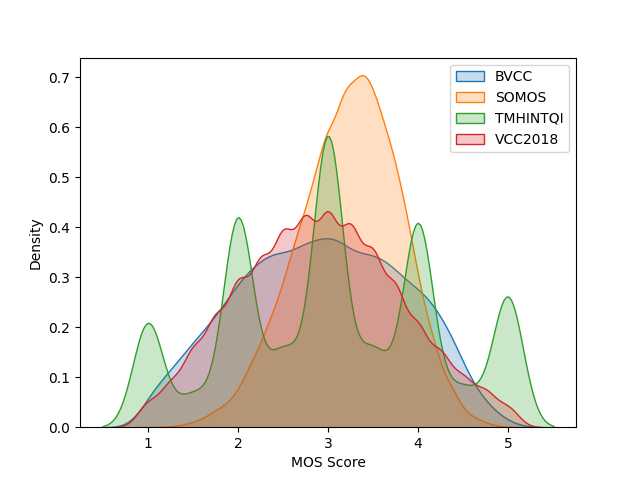}}
\end{minipage}
\caption{Data Distribution for various data sets}
\label{fig:datadistribution}
\end{figure}

We used open source data sets like \textbf{BVCC}\cite{b20}, \textbf{VCC2018}\cite{b21}, \textbf{SOMOS}\cite{b22}, \textbf{TMHINTQI}\cite{b23}. Refer Table~\ref{table:tab2} for the number of Audio Samples and Total Audio Duration (in hrs). We down sampled all audios to 16kHz.  Fig.\ref{fig:datadistribution} shows distribution of number of audio samples for the data sets used. It shows that BVCC data is normally distributed around MOS score of 3 while TMHINTQI has a varied distribution for each of the MOS scores. This varied normal distribution of TMHINTQI is the reason of poor MOS score of LE-SSL-MOS\cite{b28}.

\begin{table}[]
\caption{Details of the Dataset used for the experiment}
\resizebox{0.49\textwidth}{!}{%
\begin{tabular}{|l|l|l|l|l|l|}
\hline
\textbf{Dataset} & \textbf{Total Samples} &  \textbf{Total Duration} \\ \hline
BVCC     & 7106             & 8.02              \\ \hline
VCC 2018 & 20871            & 19.09              \\ \hline
SOMOS    & 20100            & 26.17              \\ \hline
TMHINTQI & 14915            & 13.07              \\ \hline
\end{tabular}%
}
\label{table:tab2}
\end{table}

\subsection{\textbf{Features}}
\begin{figure}[htb]
\begin{minipage}[b]{\linewidth}
    
  \centering
  \centerline{\includegraphics[width=8.5cm]{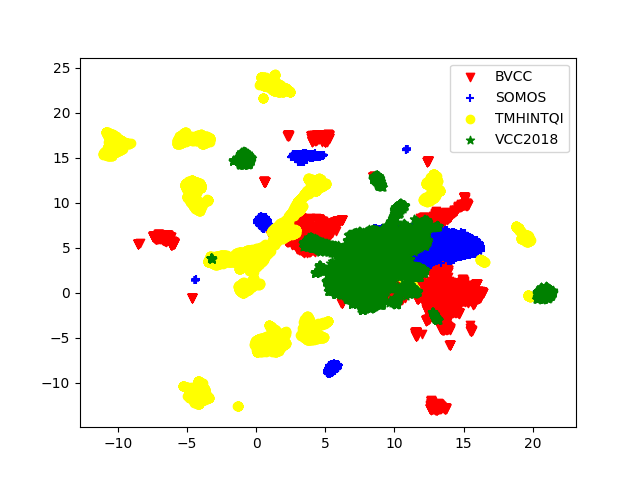}}
\end{minipage}
\caption{Wav2Vec Features Distribution}
\label{fig:wav2vecfeatures}
\end{figure}

The UMAP-based dimesionality reduction is applied on wav2vec features to get 2-D representation and plotted in Fig.~\ref{fig:wav2vecfeatures}. From Fig.~\ref{fig:wav2vecfeatures} it can be observed that even though the features are vaguely distributed in different dimensions for each of the data sets but the proposed model is able to capture audio features across different speaker and generalizes on data sets. Further it can be observed from Fig.~\ref{fig:wav2vecfeatures} that audio representation of TMHINTQI is distributed into different clusters which are far from each other while the audio representations of VCC2018 is majorly distributed in one major cluster.

\subsection{\textbf{Evaluation Metrics}}
The model evaluation is done based on 4 metrics - MSE, LCC, SRCC and KTAU.

Mean Squared Error (MSE) is used to calculate the Error for calculating MOS.
\begin{equation}
    MSE=\frac{1}{n}\sum (X_i-Y_i)^2
\end{equation}
where $n$ are the total number of observations, and $X$, $Y$ are the actual and the predicted values.

Linear Concordance Correlation Coefficient (LCC) is helpful in comparing bi-variate data.
\begin{equation}
    LCC=\frac{n(\sum xy)-(\sum x)(\sum y)}{\sqrt{[n\sum x^2 -(\sum x)^2][n\sum y^2-(\sum y)^2]}}
\end{equation}
where $x$, $y$ are the actual and the observed values respectively. 

Spearman Rank Correlation Coefficient (SRCC) describes how relationship between two variables can be assessed using monotonic function.
\begin{equation}
    SRCC=1-\frac{6\sum (x_i^2-y_i^2)}{n(n^2-1)}
\end{equation}
Where $x_i^2$ is the $i^{th}$ rank actual value, $y_i^2$ is the $i_{th}$ rank predicted value and $n$ is the total number of observations.

Kendall Rank Correlation Coefficient (KTAU) is a non-parametric relationship between columns of ranked data.
\begin{equation}
    KTAU=\frac{C - D}{C + D}
\end{equation}
where C is the number of concordant pairs and D is the number of Discordant pairs.

These metrics have been used across all works in MOS Prediction Works.

\subsection{\textbf{Observations}}
We experimented our model with 4 different inputs. We used MFCC and LFCC as 2-dimensional data for generating the latent representations. We also experimented with Speaker Embeddings like x-vector~\cite{b29}. To capture more information about the audio, we used SSL feature (wav2vec) and trained our model with just 1574 model parameters. See Table~\ref{table:results} where we achieved SOTA results and beat all the Research work done in MOS prediction.

\begin{figure}[htb]
\begin{minipage}[b]{\linewidth}
    
  \centering
  \centerline{\includegraphics[width=8.5cm]{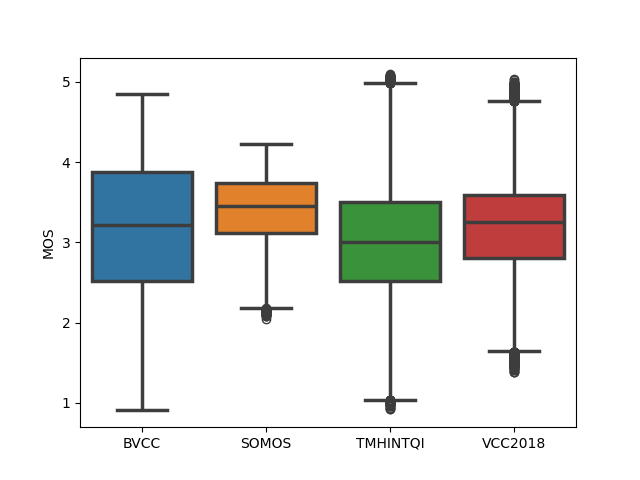}}
\end{minipage}
\caption{Box Plot for MOS on different data sets}
\label{fig:box_plot}
\end{figure}

It can be seen in Fig-\ref{fig:box_plot} that our model generalizes well on the data sets used and the MOS Score for the test data are evenly distributed. This means that our model is not biased towards predicting MOS score of more than 4 or MOS Score of less than 2.

\subsection{\textbf{Ablation Study}}
We would like to share more insights for feature selection and depth for our model architecture. Although SSL feature like wav2vec suits the best for our model architecture but we also experimented with other feature like MFCC, LFCC and X-Vector. For these features our system was able to beat few old systems like MOSNet and MBNet but was not able to generalize well on the data sets used. 

\begin{table}[]
\caption{Feature Experimentation on BVCC}
\resizebox{0.49\textwidth}{!}{%
\begin{tabular}{|l|l|l|l|l|l|}
\hline
\textbf{Feature} &  \textbf{MSE} &  \textbf{LCC}  &  \textbf{SRCC} &  \textbf{KTAU} \\ \hline
MFCC             & 0.56  & 0.54          &   0.55            &   0.49               \\ \hline
LFCC             & 0.43  & 0.683          &   0.68            &   0.62              \\ \hline
X-Vector         & 0.48  & 0.47  &   0.48            &   0.43                        \\ \hline
wav2vec          & 0.144  & 0.948          &   0.946            &   0.81            \\ \hline
\end{tabular}%
}
\label{table:tab3}
\end{table}

\begin{figure}[htb]
\begin{minipage}[b]{\linewidth}
    
  \centering
  \centerline{\includegraphics[width=8.5cm]{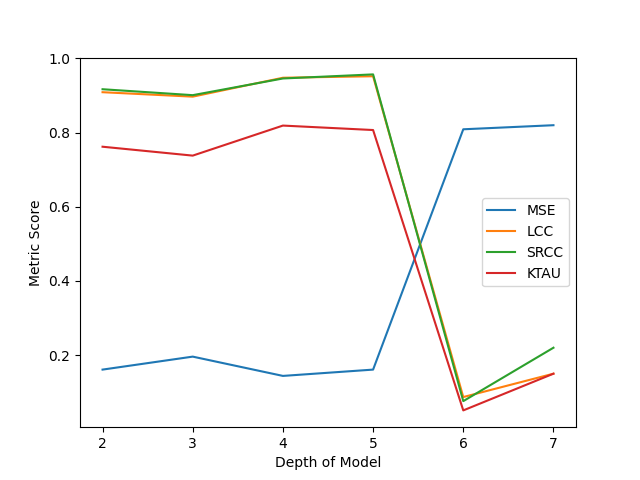}}
\end{minipage}
\caption{Model Depth effect on BVCC}
\label{fig:abalation}
\end{figure}

Also the depth of the model is also a parameter that we experimented. Fig-\ref{fig:abalation} shows the effect of depth of the model for BVCC data set on MSE, LCC, SRCC and KTAU. A Good MOS Model should achieve minimum MSE while LCC, SRCC and KTAU should be as high as possible. As the depth of the model increases, the model starts underfitting the data which results in poor MSE. We see that the optimal number of depth of model is 4 with four double convolutional layers and three down sampling layer. 

\section{\textbf{CONCLUSION}}
In this paper, we are able to show that our novel model SALF-MOS generalizes well for MOS prediction tasks and will help a lot to evaluate TTS and Voice Conversion models. It will narrow down the funds required for human evaluators and system infrastructure for human evaluators. Our small size model requires very less amount of resources to run and its end-to-end capabilities and non-reliability on training SSL model makes it easier to setup. Even without pre-training, our model outperforms all state-of-the-art results and sets a new benchmark. Our model will help to reduce subjective dependency to measure the quality of speech samples.

\bibliographystyle{IEEEtran}
\bibliography{SPCOM_cameraready}
\end{document}